# Some minimal characteristics and flat trajectories of a system in the $N$-body problem

Igor Pavlov

Abstract: For the classical model of a system of $N$ particles with interaction (the $N$-body problem) we consider some spatial problems of nonlinear programming for finding conditional minima and local extrema for the main characteristics of the system on its phase space. We prove that these conditional minima and local extrema of the system characteristics can only be achieved on flat, in general, not necessarily finite, trajectories of the system's motion in space, (under the condition that the system angular momentum is non-zero). And then we find the basic conditions limiting the values of the integrals of motion: angular momentum and total energy of the system under which these trajectories are flat finite ("circular" or elliptical) trajectories on which the flat structure of the system is preserved in time under the certain restrictions on the values of the system's integrals of motion. And we determine also the values of the system's key characteristics on these flat trajectories. Next, in the Appendix, we consider some additional related results for a more general model of the system.

## 1. Introduction

Consider a system of $N$ different material points (particles). Let $\boldsymbol{r} = (\boldsymbol{r}_1, \boldsymbol{r}_2, \dots, \boldsymbol{r}_N) \in \mathfrak{R}$ be the complete set of radius-vectors of all particles of the system in three-dimensional space, $\boldsymbol{r}_i = (x_i, y_i, z_i)$ be the radius-vector of the particle with index $i$; where the coordinate space $\mathfrak{R} = \{\boldsymbol{r} : |\boldsymbol{r}_j - \boldsymbol{r}_i| > 0, \ i \neq j \} \subset E_{3N}$. (Here and below, indexes $i, j$ take the values 1, …, $N$; vectors are denoted in bold, $E_n$ is the $n$-dimensional Euclidean space.) We also denote the set of velocity vectors of various particles of the system as $\boldsymbol{v} = (\boldsymbol{v}_1, \boldsymbol{v}_2, \dots, \boldsymbol{v}_N)$, where $\boldsymbol{v}_i = (\propto_i, \beta_i, \gamma_i)$ is the velocity vector of the $i$ particle; $\boldsymbol{v} \in \mho$, where $\mho$ is the set of possible values of $\boldsymbol{v} = (\boldsymbol{v}_1, \boldsymbol{v}_2, \dots, \boldsymbol{v}_N)$, $\mho = E_{3N}$. The set $\Omega$ (phase space) of all possible states $(\boldsymbol{r}, \boldsymbol{v})$ of the system in this model is an open subset of the $6N$-dimensional Euclidean space of the form: $\Omega = \{(\boldsymbol{r}, \boldsymbol{v}) \in E_{6N}: |\boldsymbol{r}_j - \boldsymbol{r}_i| > 0, \ i \neq j\} \subset E_{6N}$.

The trajectory of the system $[\boldsymbol{r}(t), \boldsymbol{v}(t)]$ is determined by a system of differential equations with some initial conditions:

$$m_i \dot{\boldsymbol{v}}_i = \boldsymbol{F}_i(\boldsymbol{r}), \quad i = 1, \dots, N \tag{1.1}$$

$$\dot{\boldsymbol{r}}_i = \boldsymbol{v}_i, \quad i = 1, \dots, N \tag{1.2}$$

Where the interaction force $\boldsymbol{F}_i(\boldsymbol{r})$ acting on the particle $i$ from all other particles of the system has the form:

$$\boldsymbol{F}_i(\boldsymbol{r}) = \sum_{j \neq i} \boldsymbol{F}_{ij}(\boldsymbol{r}), \qquad \boldsymbol{F}_{ij}(\boldsymbol{r}) = \frac{\gamma \, m_i \, m_j \, \boldsymbol{L}_{ij}(\boldsymbol{r})}{|\boldsymbol{L}_{ij}(\boldsymbol{r})|^3}, \quad i \neq j$$

where $\boldsymbol{F}_{ij}(\boldsymbol{r})$ is the interaction force acting on particle $i$ from particle j, $m_i > 0$ is the mass of particle $i$; $\gamma > 0$ is some constant, $\boldsymbol{L}_{ij}(\boldsymbol{r}) = \boldsymbol{r}_j - \boldsymbol{r}_i$.This model corresponds to the known classical N–body problem, where the relations for $\boldsymbol{F}_{ij}(\boldsymbol{r})$ correspond to Newton's law of gravity, when the force of attraction between two particles with masses $m_i, m_j$ is inversely proportional the square of the distance between them.

Further we assume that the origin is in the center of mass of the system. A trajectory $[\boldsymbol{r}(t), \boldsymbol{v}(t)]$ satisfying the system of differential equations (1.1), (1.2) under some initial conditions will be called finite (in terms [1], et.al.) if there exist constants $0 < C_1 < C_2 < \infty$ such that $C_1 < |\boldsymbol{r}_j(t) - \boldsymbol{r}_i(t)| < C_2$ for each $i \neq j$, $t \geq 0$. As a characteristic of the size of the system in the state $(\boldsymbol{r}, \boldsymbol{v}) \in \Omega$ we take its "root-mean-square size" (further simply the "size") $b = b(\boldsymbol{r}) = \sqrt{\sum_i \beta_i |\boldsymbol{r}_i|^2}$, where $\beta_i = m_i / \sum_j m_j$, $i = 1, \dots, N$. To simplify the formulas below, as a characteristic of the size of the system, we'll also use the function $g = g(\boldsymbol{r}) = \sum_i m_i |\boldsymbol{r}_i|^2$, which is related to $b = b(\boldsymbol{r})$ by the relation $g = b^2 \sum_j m_j$ . (Function $g$ also has the meaning of the moment of inertia of the system in a problem on a plane.)

Next, in Section 2 we first consider a spatial problem of nonlinear programming to find local extrema of the system total energy function at fixed (non-zero) value of angular momentum of the system. Based on the solution of this problem in Theorem 2.1, we prove that these local extrema can be achieved only on flat finite circular trajectories on which the flat structure of the system is preserved in time. Further, it will follow from the results of Section 3 that along any trajectory of the system's motion in space, at any moment in time, the size of the system cannot be less than a certain lower bound (in general, unknown) which depends on the system kinetic energy at this moment and on the number N of particles and their masses. And we determine this lower bound as the solution to a certain nonlinear programming problem of dimension 6N for finding a conditional minimum of the size of the system on its phase space, where the system kinetic energy at a given moment of time is one of the problem's input parameters. Based on the solution of this problem in Theorem 3.1, we prove that the minimum size of a system (with non-zero angular momentum) at the given current level of its kinetic energy can be achieved only on flat, in general, not necessarily finite, trajectories of the system's motion in space. Next, in Section 3 we will obtain the basic conditions limiting the values of the integrals of motion: angular momentum and total energy of the system under which these trajectories are flat finite circular or elliptical trajectories, on which the flat structure of the system is preserved in time (up to rotation and scale) under the certain restrictions on the values of the system's integrals of motion. And these conditions are largely related to the solution of the previous spatial problem in Section 2 for finding local extrema of the total energy function of the system. A similar result is also hold, namely: on these same flat finite trajectories, the minimum possible values of the kinetic energy and potential function of the system are achieved at the given current value of its size. In Section 3 we find also the values of the system's key characteristics on these flat trajectories. Next, in the Appendix we prove some related additional results for a more general model of the system.

## 2. Local extrema of the total energy function of the system

Let $\boldsymbol{L}(\boldsymbol{r}, \boldsymbol{v}) = \sum_i [\boldsymbol{r}_i, m_i \boldsymbol{v}_i]$ be the angular momentum of the system and $T(\boldsymbol{v}) = \sum_i m_i |\boldsymbol{v}_i|^2/2$ is the kinetic energy of the system in the state $(\boldsymbol{r}, \boldsymbol{v}) \in \Omega$. Let's denote by

$$f(\boldsymbol{r}) = (1/2) \sum_i \sum_{j \neq i} \frac{\gamma\, m_i\, m_j}{|\boldsymbol{r}_j - \boldsymbol{r}_i|}\,, \qquad \boldsymbol{r} \in \Re$$

the potential (or "force") function, which coincides, up to a sign and an arbitrary constant, with the potential energy $U(\boldsymbol{r})$ of the system: $f(\boldsymbol{r}) = -U(\boldsymbol{r}) + C$ [2], [3]. And let $E(\boldsymbol{r}, \boldsymbol{v}) = T(\boldsymbol{v}) - f(\boldsymbol{r})$ is the total energy function of the system in state $(\boldsymbol{r}, \boldsymbol{v}) \in \Omega$. Next, we first consider the spatial problem of finding local extrema of the function $E = E(\boldsymbol{r}, \boldsymbol{v})$ of the total energy of the system on its phase space at fixed value of the system angular momentum, i.e. under the conditions: $\boldsymbol{L}(\boldsymbol{r}, \boldsymbol{v}) = \boldsymbol{L}$, $(\boldsymbol{r}, \boldsymbol{v}) \in \Omega$, where $\boldsymbol{L} = (L_x, L_y, L_z) \neq \boldsymbol{0}$ is some constant. Condition at the angular momentum of the system corresponds to three conditions on the components of this vector. We further assume without loss of generality that the angular momentum vector of the system is directed along the axis $z$, i.e. $\boldsymbol{L} = (0, 0, L)$, where $L = |\boldsymbol{L}| > 0$. In more detail, this problem in variables $(x_i, y_i, z_i)$, $(\propto_i, \beta_i, \gamma_i)$, $i = 1, \ldots, N$; has the following form: It is required to find

$$\text{local extrema } E(\boldsymbol{r}, \boldsymbol{v}), \tag{2.1}$$

where the local extrema of the function $E(\boldsymbol{r}, \boldsymbol{v})$ are calculated over all states $(\boldsymbol{r}, \boldsymbol{v}) \in \Omega$ under the conditions:

$$L_x(\boldsymbol{r}, \boldsymbol{v}) = \sum_i m_i (y_i \gamma_i - z_i \beta_i) = 0 \tag{2.2}$$

$$L_y(\boldsymbol{r}, \boldsymbol{v}) = \sum_i m_i (z_i \propto_i - x_i \gamma_i) = 0 \tag{2.3}$$

$$L_z(\boldsymbol{r}, \boldsymbol{v}) = \sum_i m_i (x_i \beta_i - y_i \propto_i) = L \tag{2.4}$$

where constant $L > 0$ is the parameter of this problem. It's not difficult to show that the global minimum and global maximum of the function $E(\boldsymbol{r}, \boldsymbol{v})$ under the conditions (2.2) – (2.4) are not achieved: $\inf E(\boldsymbol{r}, \boldsymbol{v}) = -\infty$, $\sup E(\boldsymbol{r}, \boldsymbol{v}) = \infty$, if $N > 2$. Next, we will prove that local extrema of the system total energy function $E(\boldsymbol{r}, \boldsymbol{v})$ under the conditions (2.2) – (2.4) can only be achieved on flat finite (circular) trajectories. Let's denote by $\mathfrak{D}(L) \subset \Omega$ the subset of states $(\boldsymbol{r}, \boldsymbol{v}) \in \Omega$ of the phase space for which conditions (2.2) – (2.4) are satisfied.

**Theorem 2.1** *Let at some point $(\boldsymbol{r}, \boldsymbol{v}) \in \mathfrak{D}(L) \subset \Omega$ there is a local extremum of the function $E(\boldsymbol{r}, \boldsymbol{v})$ on the set $\mathfrak{D}(L) \subset \Omega$ (where the parameter $L > 0$). Then at this point the equalities $z_i = 0$, $\gamma_i = 0$ are satisfied for each $i = 1, \ldots, N$.*

*Proof.* At the problem (2.1) – (2.4) all functions $E$, $L_x, L_y, L_z$ are continuous and have continuous partial derivatives at each point $(\boldsymbol{r}, \boldsymbol{v}) \in \Omega$. Necessary conditions [4] – [6] of a local extremum of the function $E = E(\boldsymbol{r}, \boldsymbol{v})$ under the conditions (2.2) – (2.4) at point $(\boldsymbol{r}, \boldsymbol{v}) \in \mathfrak{D}(L) \subset \Omega$ have the form:

$$\nabla E = a_1 \nabla L_x + a_2 \nabla L_y + a_3 \nabla L_z \tag{2.5}$$

where $a_1$, $a_2$, $a_3$ are some coefficients (Lagrange multipliers). Note that gradient vectors on the right side of (2.5) are linearly independent at any point $(\boldsymbol{r}, \boldsymbol{v}) \in \mathfrak{D}(L)$ due to the condition $L > 0$ (see below Remark 2.1), which ensures the correctness of the necessary minimum condition in (2.5). Whence, taking into account that $\left({\partial f}/{\partial \boldsymbol{v}_i}\right) = \boldsymbol{0}$, $\left({\partial T}/{\partial \boldsymbol{r}_i}\right) = \boldsymbol{0}$, $i = 1, \ldots, N$; the equations follow:

$$-\frac{\partial f}{\partial \boldsymbol{r}_i} = a_1 \frac{\partial L_x}{\partial \boldsymbol{r}_i} + a_2 \frac{\partial L_y}{\partial \boldsymbol{r}_i} + a_3 \frac{\partial L_z}{\partial \boldsymbol{r}_i}, \quad i = 1, \ldots, N \tag{2.6}$$

$$\frac{\partial T}{\partial \boldsymbol{v}_i} = a_1 \frac{\partial L_x}{\partial \boldsymbol{v}_i} + a_2 \frac{\partial L_y}{\partial \boldsymbol{v}_i} + a_3 \frac{\partial L_z}{\partial \boldsymbol{v}_i}, \quad i = 1, \ldots, N \tag{2.7}$$

Equations (2.6) in the variables $(x_i, y_i, z_i)$, $(\propto_i, \beta_i, \gamma_i)$, taking into account that $\left({\partial f}/{\partial \boldsymbol{r}_i}\right) = \boldsymbol{F}_i(\boldsymbol{r})$, have the following form:

$$-F_{ix}(\boldsymbol{r}) = m_i(-a_2 \gamma_i + a_3 \beta_i), \quad i = 1, \ldots, N \tag{2.8}$$

$$-F_{iy}(\boldsymbol{r}) = m_i(a_1\gamma_i - a_3 \propto_i), \quad i = 1,\dots,N \tag{2.9}$$

$$-F_{iz}(\boldsymbol{r}) = m_i(-a_1\beta_i + a_2 \propto_i), \quad i = 1,\dots,N \tag{2.10}$$

where $F_{ix}(\boldsymbol{r})$, $F_{iy}(\boldsymbol{r})$, $F_{iz}(\boldsymbol{r})$ are the coordinates of vector $\boldsymbol{F}_i(\boldsymbol{r})$. Equations (2.7) in the variables $(x_i, y_i, z_i)$, $(\propto_i, \beta_i, \gamma_i)$ are written as

$$\propto_i = a_2 z_i - a_3 y_i\,, \quad i = 1,\dots,N \tag{2.11}$$

$$\beta_i = -a_1 z_i + a_3 x_i\,, \quad i = 1,\dots,N \tag{2.12}$$

$$\gamma_i = a_1 y_i - a_2 x_i\,, \quad i = 1,\dots,N \tag{2.13}$$

Next, we introduce the vector of coefficients $\mathbf{a} = (a_1, a_2, a_3)$. Then these equations can be written in a more compact vector form:

$$-\boldsymbol{F}_i(\boldsymbol{r}) = m_i[\boldsymbol{v}_i\mathbf{a}], \quad \boldsymbol{v}_i = [\mathbf{a}\,\boldsymbol{r}_i], \quad i = 1,\dots,N \tag{2.14}$$

where [ · ] is the vector multiplication. Or in matrix form:

$$-\boldsymbol{F}_i(\boldsymbol{r}) = m_i\|A\|\,\boldsymbol{v}_i\,, \ -\boldsymbol{v}_i = \|A\|\,\boldsymbol{r}_i\,, \quad i = 1,\dots,N \tag{2.15}$$

where the matrix $\|A\|$ has the form:

$$\|A\| = \begin{Vmatrix} 0 & a_3 & -a_2 \\ -a_3 & 0 & a_1 \\ a_2 & -a_1 & 0 \end{Vmatrix}$$

From equations (2.14) it follows, taking into account the condition $L > 0$, that at any point of local extremum $(\boldsymbol{r}, \boldsymbol{v}, a_1, a_2, a_3), (\boldsymbol{r}, \boldsymbol{v}) \in \mathfrak{D}(L)$ of the function E on the set $\mathfrak{D}(L) \subset \Omega$ the vector of coefficients $\mathbf{a} = (a_1, a_2, a_3) \neq \mathbf{0}$. And all particles of the system (except, may be, one central particle) at any point of local extremum have non-zero vectors $\boldsymbol{r}_i$, $\boldsymbol{v}_i$ and the velocity vector $\boldsymbol{v}_i$ of each particle is perpendicular to its radius vector $\boldsymbol{r}_i$.

Let's denote by $\mathbf{p_1} = (0, -a_3, a_2)$, $\mathbf{p_2} = (a_3, 0, -a_1)$, $\mathbf{p_3} = (-a_2, a_1, 0)$ the column vectors of the matrix $\|A\|$. Then, by virtue of (2.15), at point of local extremum $(\boldsymbol{r}, \boldsymbol{v}, a_1, a_2, a_3)$, $(\boldsymbol{r}, \boldsymbol{v}) \in \mathfrak{D}(L)$ of the function $E$ on the set $\mathfrak{D}(L) \subset \Omega$ the following equalities must hold:

$$\boldsymbol{F}_i(\boldsymbol{r}) = -m_i(\propto_i \mathbf{p_1} + \beta_i\,\mathbf{p_2} + \gamma_i\mathbf{p_3})\,, \quad i = 1,\dots,N \tag{2.16}$$

$$\boldsymbol{v}_i = -(x_i\,\mathbf{p_1} + y_i\,\mathbf{p_2} + z_i\mathbf{p_3})\,, \quad i = 1,\dots,N \tag{2.17}$$

where $\mathbf{a} = (a_1, a_2, a_3) \neq \mathbf{0}$. The determinant of the matrix $\|A\|$ is equal to zero and the vectors $\mathbf{p_1}$, $\mathbf{p_2}$, $\mathbf{p_3}$ are coplanar (in particular, $a_1\mathbf{p_1} + a_2\mathbf{p_2} + a_3\mathbf{p_3} = \mathbf{0}$) and these vectors, evidently, cannot be collinear. That is, by virtue of (2.16), (2.17), at any point of local extremum of the function $E$ on the set $\mathfrak{D}(L) \subset \Omega$ all vectors $\boldsymbol{F}_i(\boldsymbol{r})$, $\boldsymbol{v}_i$ are in the same plane spanned by coplanar vectors $\mathbf{p_1}, \mathbf{p_2}, \mathbf{p_3}$. It is not difficult to further show that if all vectors $\boldsymbol{F}_i(\boldsymbol{r})$, $i = 1,\dots,N$ are in the same plane, then all radius-vectors $\boldsymbol{r}_i$, $i = 1,\dots,N$ are also in this plane. Thus, at any point of local extremum of the function $E$ on the set $\mathfrak{D}(L) \subset \Omega$ all vectors $\boldsymbol{r}_i, \boldsymbol{v}_i$ are in the same plane spanned by coplanar vectors $\mathbf{p_1}$, $\mathbf{p_2}$, $\mathbf{p_3}$. Due to the definition of the angular momentum of the system, this plane is perpendicular to the angular momentum vector $\boldsymbol{L} = (0, 0, L)$ and the following equations are fulfilled: $(\mathbf{p_1}, \boldsymbol{L}) = a_2 L = 0$, $(\mathbf{p_2}, \boldsymbol{L}) = -a_1 L = 0$, $(\mathbf{p_3}, \boldsymbol{L}) = 0$. Whence it follows that the coefficients $a_1 = a_2 = 0$, which corresponds to the transition of the above spatial problem of nonlinear programming in (2.1) – (2.4) into a problem of nonlinear programming on a plane (perpendicular to the angular momentum vector $\boldsymbol{L}$). And since the angular momentum vector is directed along the axis z, this plane can only be the plane $z = 0$, which proves Theorem 2.1. ■

**Remark 2.1** It's not difficult to show that gradient vectors on the right side of (2.5) are linearly independent at any point $(\boldsymbol{r}, \boldsymbol{v}) \in \mathfrak{D}(L)$ due to the condition $L > 0$. Indeed, if these vectors are linearly dependent at some point $(\boldsymbol{r}, \boldsymbol{v}) \in \mathfrak{D}(L)$, then at this point the right-hand side of (2.5) is equal to zero for some vector of coefficients $\mathbf{a} = (a_1, a_2, a_3) \neq \mathbf{0}$. From this, similar to (2.6) – (2.14), the equations follow: $[\boldsymbol{v}_i \mathbf{a}] = \mathbf{0}$, $[\boldsymbol{r}_i \mathbf{a}] = \mathbf{0}$, $i = 1, \dots, N$. But from these equations it follows, that all non-zero vectors $\boldsymbol{r}_i$, $\boldsymbol{v}_i$ are collinear, which contradicts the condition $L > 0$.

Next, based on Theorem 2.1, we prove that the local extrema of the system total energy function E at fixed value of the system angular momentum $L = |\boldsymbol{L}| > 0$ can only be achieved on flat finite circular trajectories. Let at some point of the phase space $(\boldsymbol{r}^*, \boldsymbol{v}^*) \in \mathfrak{D}(L) \subset \Omega$ (where $L > 0$) a local extremum of the function $E$ on the set $\mathfrak{D}(L)$ is achieved. Then a point $(\boldsymbol{r}^*, \boldsymbol{v}^*, a_1, a_2, a_3)$ is a point of local extremum of this function in the spatial problem (2.1) – (2.4), where the vector of coefficients $\mathbf{a} = (a_1, a_2, a_3) \neq \mathbf{0}$. And by virtue of Theorem 2.1, at this point all vectors $\boldsymbol{r}_i^*, \boldsymbol{v}_i^*$, $i = 1, \dots, N$ are in the same plane $z = 0$, and the coefficients $a_1 = a_2 = 0$. Conditions (2.8) – (2.13) for the coordinates of vectors $\boldsymbol{r}_i^* = (x_i^*, y_i^*)$, $\boldsymbol{v}_i^* = (\propto_i^*, \beta_i^*)$ on this plane take the following form:

$$F_{ix}(\boldsymbol{r}^*) = -m_i a_3 \beta_i^*, \quad F_{iy}(\boldsymbol{r}^*) = m_i a_3 \propto_i^*, \quad i = 1, \dots, N$$

$$\propto_i^* = -a_3 y_i^*, \quad \beta_i^* = a_3 x_i^*, \quad i = 1, \dots, N$$

where $a_3 \neq 0$. From where the equalities follow:

$$\boldsymbol{F}_i(\boldsymbol{r}^*) = -m_i \omega^2 \boldsymbol{r}_i^*, \quad i = 1, \dots, N \tag{2.18}$$

$$\propto_i^* = -\omega y_i^*, \quad \beta_i^* = \omega x_i^*, \quad i = 1, \dots, N \tag{2.19}$$

where $\omega = a_3$. Relations (2.18), (2.19) define a flat periodic "circular" trajectory [7], on which the particles of the system rotate around the origin: center of mass of the system with an angular velocity ω and on which the flat structure of the system is preserved in time (up to rotation). Next, we will determine the value of ω and other characteristics of the system on this trajectory depending on the value of the parameter $L > 0$.

**Remark 2.2** It also follows from these relations that at any point $(\boldsymbol{r}^*, \boldsymbol{v}^*) \in \mathfrak{D}(L)$ of the local extremum of the total energy function $E(\boldsymbol{r}, \boldsymbol{v})$ on the set $\mathfrak{D}(L) \subset \Omega$ the following equalities are satisfied: $\sum_i m_i \boldsymbol{r}_i^* = \mathbf{0}$, $\sum_i m_i \boldsymbol{v}_i^* = \mathbf{0}$.

**Remark 2.3** Note that the functions on the right-hand side of the system of the above differential equations (1.1), (1.2) are continuous and have continuous partial derivatives at each point $(\boldsymbol{r}, \boldsymbol{v})$ of the open set $\Omega \subset E_{6N}$, i.e. these functions satisfy the known conditions ([8] et. al.) of existence and uniqueness of a solution (trajectory), starting at some moment $t_0$ from the initial state $(\boldsymbol{r}, \boldsymbol{v}) \in \Omega$, on some interval $(\tau_1, \tau_2)$ such that $-\infty \leq \tau_1 < t_0 < \tau_2 \leq \infty$. From this, taking into account the above Theorem 2.1, it follows that the trajectory of the system, starting at some moment of time $t_0$ from any point $(\boldsymbol{r}^*, \boldsymbol{v}^*) \in \mathfrak{D}(L) \subset \Omega$ of local extremum of the function $E = E(\boldsymbol{r}, \boldsymbol{v})$ on the set $\mathfrak{D}(L) \subset \Omega$, is flat (in the plane $z = 0$).

Thus, from Theorem 2.1 it follows that the local extrema of the total energy function $E$ at fixed value of the system angular momentum $L = |\boldsymbol{L}| > 0$ can only be achieved on the above flat finite circular trajectories (on which the flat structure of the system is preserved in time).

Next, we will find the values of the total energy $E$ and other characteristics of the system on these trajectories depending on the value of the parameter $L > 0$.

**The values of the system characteristics.** Let $\boldsymbol{r} = (\boldsymbol{r}_1, \boldsymbol{r}_2, \dots, \boldsymbol{r}_N) \in \mathfrak{R}_2$ be the complete set of radius vectors of all particles of the system on the plane $z = 0$, where $\boldsymbol{r}_i = (x_i, y_i)$ is the radius vector of the particle $i$ and $\mathfrak{R}_2 = \{\boldsymbol{r} : |\boldsymbol{r}_j - \boldsymbol{r}_i| > 0,\ i \neq j\} \subset E_{2N}$ is the coordinate space of the system on this plane. Consider the following problem of nonlinear programming: It is required to find

$$min\,[f(\boldsymbol{r}) + \lambda g(\boldsymbol{r})], \quad \boldsymbol{r} \in \mathfrak{R}_2 \tag{2.20}$$

where the minimum is calculated over all states $\boldsymbol{r} \in \mathfrak{R}_2$. It's not difficult to show ([7], par.2) that global minimum in (2.20) is achieved (for any $\lambda > 0,\ N > 1,\ m_i > 0,\ i = 1, \dots, N$) at some, not necessarily unique, point $\boldsymbol{r}_\lambda = (\boldsymbol{r}_{\lambda 1}, \ . \ . \ . \ , \boldsymbol{r}_{\lambda N}) \in \mathfrak{R}_2$. And at this point the following equalities (necessary conditions for a minimum) are fulfilled:

$$\boldsymbol{F}_i(\boldsymbol{r}_\lambda) = -2\lambda m_i \boldsymbol{r}_{\lambda i}, \quad i = 1, \dots, N \tag{2.21}$$

The equalities (2.21) define a flat periodic circular trajectory ([7], par.3) of the form $\boldsymbol{r}_\lambda(t) \in \mathfrak{R}_2,\ t \geq 0$ (on which the particles of the system rotate around the origin: center of mass with an angular velocity $\sqrt{2\lambda}$), on which the flat structure of the system is preserved in time (up to rotation). The main characteristics of the system on this trajectory are constant and have the following form ([7], par.4): $g = (C/2\lambda)^{2/3},\ f = C/\sqrt{g},\ T = f/2,$

$$E = T - f = -(C/2)^{2/3}\,\lambda^{1/3}, \quad L = \sqrt{2\lambda}g = C^{2/3}/(2\lambda)^{1/6}$$

where $C = C(\boldsymbol{m}) > 0$ is a function of the masses $\boldsymbol{m} = (m_1, m_2, \dots, m_N)$ of the particles of the system, corresponding to the global minimum in (2.20), for which the following lower estimate ([7], par.5) is fulfilled (for any $N > 1, m_i > 0,\ i = 1, \dots, N$):

$$C(\boldsymbol{m}) \geq \gamma\left(\sum_{i<j} m_i m_j\right)^{3/2} / \left(\sum_j m_j\right)^{1/2} \tag{2.22}$$

And this lower estimate gives the exact value of the function $C(\boldsymbol{m})$ for cases $N = 2$ and $N = 3$ ([7], par.6). Equalities (2.18) and (2.21) coincide if $\omega = \sqrt{2\lambda}$. Expressing the above characteristics through the parameter $L > 0$ from the previous formulas, we find:

$$\omega = [C(\boldsymbol{m})]^2/L^3, \quad E = -[C(\boldsymbol{m})]^2/(2L^2) < 0 \tag{2.23}$$

$$g = L^4/[C(\boldsymbol{m})]^2,\ f = [C(\boldsymbol{m})]^2/L^2,\ T = [C(\boldsymbol{m})]^2/(2L^2) \tag{2.24}$$

Note that relations of the form (2.21) are fulfilled not only for the above point $\boldsymbol{r}_\lambda = (\boldsymbol{r}_{\lambda 1}, \ . \ . \ . \ , \boldsymbol{r}_{\lambda N}) \in \mathfrak{R}_2$ of the global minimum in problem (2.20), but also for the point $\boldsymbol{r}_\lambda^{(k)} = \left(\boldsymbol{r}_{\lambda 1}^{(k)}, \ . \ . \ . \ , \boldsymbol{r}_{\lambda N}^{(k)}\right) \in \mathfrak{R}_2$ of any kth local minimum in this problem, namely:

$$\boldsymbol{F}_i\left(\boldsymbol{r}_\lambda^{(k)}\right) = -2\lambda m_i \boldsymbol{r}_{\lambda i}^{(k)}, \quad i = 1, \dots, N$$

According to [7], par.7 these equalities define, similarly to equalities (2.21), a flat periodic circular trajectory of the form $\boldsymbol{r}_\lambda^{(k)}(t) \in \mathfrak{R}_2,\ t \geq 0$ (on which the flat structure of the system is also preserved in time) with the following system characteristics on this trajectory:

$$g = (C_k/2\lambda)^{2/3},\ f = C_k/\sqrt{g},\ T = f/2,$$

$$E = T - f = -(C_k/2)^{2/3}\,\lambda^{1/3},\ L = \sqrt{2\lambda}g = {C_k}^{2/3}/(2\lambda)^{1/6}$$

where $C_k = C_k(\boldsymbol{m})$ is a function of the masses $\boldsymbol{m} = (m_1, m_2, \dots, m_N)$ of the particles, corresponding to the kth local minimum, for which the following lower estimate holds:

$$C_k(\boldsymbol{m}) \geq C(\boldsymbol{m}) \geq \gamma\left(\sum_{i<j} m_i m_j\right)^{3/2} / \left(\sum_j m_j\right)^{1/2}, \; k = 1, \dots, n; \qquad (2.25)$$

where $n$ is the number of local minima (up to equivalence relations [7], par.7) in the problem (2.20). We formally assume in these relations and further: $C_1(\boldsymbol{m}) = C(\boldsymbol{m})$. Repeating the previous relations for the case of a global minimum in (2.20), we find that the local extrema of the total energy function $E(\boldsymbol{r}, \boldsymbol{v})$ at a fixed value of $L > 0$ can only be achieved on flat finite circular trajectories, generated by the flat central configurations in (2.18), (2.19) with parameters $\omega_k = [C_k(\boldsymbol{m})]^2 / L^3$, $k = 1, \dots, n$. And the values of the system total energy on these trajectories have the form:

$$E_k = -[C_k(\boldsymbol{m})]^2 / (2L^2) < 0, \;\; k = 1, \dots, n \qquad (2.26)$$

Other characteristics of the system on these trajectories take the following values:

$$g_k = L^4 / [C_k(\boldsymbol{m})]^2, \; f_k = [C_k(\boldsymbol{m})]^2 / L^2, \; T_k = [C_k(\boldsymbol{m})]^2 / (2L^2), \; k = 1, \dots, n. \quad (2.27)$$

Thus, all local extrema $E_k$ of the total energy function of the system (with non-zero angular momentum $L > 0$) are always negative, which was not obvious in advance, and are achieved on the above flat finite circular trajectories (on which the flat structure of the system is preserved in time).

**Remark 2.4** On these same flat finite trajectories, according to the following Section 3, the minimum value of the size of the system (with non-zero angular momentum) is achieved at the current level of its kinetic energy. And the minimum values of the functions $T, f$ of the kinetic and potential energy of the system at the current value of its size are simultaneously achieved (see also Remark 3.3 and Remark 3.7 below).

**Remark 2.5** The previous relations depend on the functions $C(\boldsymbol{m}), C_k(\boldsymbol{m})$ which in the general case, if $N > 3$, are quite complex. However, using these functions allows us to determine essential relationships for the characteristics of the system (see also the following Section 3). The lower bounds for these functions on the right-hand side of (2.22), (2.25) also allows us to find the estimates for the system characteristics depending on the parameters $N, m_1, m_2, \dots, m_N$.

Next, we will consider some spatial problems of nonlinear programming, close in physical meaning, for finding conditional minima of the characteristics of the system, which can also be achieved only on flat, in general, not necessarily finite, trajectories (provided that the system angular momentum is non-zero). And under the conditions limiting the values of the integrals of motion: angular momentum and total energy of the system, these trajectories are flat finite (circular or elliptical) trajectories, on which the flat structure of the system is preserved in time under the restrictions on the values of the system's integrals of motion.

## 3. Minimum size and flat trajectories of the system

Consider the following spatial problem of nonlinear programming for finding a minimum: It is required to find

$$\min \; g(\boldsymbol{r}) \qquad (3.1)$$

where the minimum is taken over all states $(\boldsymbol{r}, \boldsymbol{v}) \in \Omega$ under the conditions:

$$\boldsymbol{L}(\boldsymbol{r}, \boldsymbol{v}) = \boldsymbol{L}, \;\; T(\boldsymbol{v}) = T \qquad (3.2)$$

Where $\boldsymbol{L} = (L_x,\ L_y,\ L_z\ ) \neq \boldsymbol{0}$, $T > 0$ are some constants. Condition at the angular momentum of the system corresponds to three conditions on the components of this vector. Just as above, we will assume without loss of generality that the angular momentum vector of the system is directed along the axis $z$, i.e. $\boldsymbol{L} = (0, 0,\ L\ )$, where $L = |\boldsymbol{L}| > 0$. In more detail, this problem in variables $(x_i\ , y_i, z_i)$, $(\propto_i\ , \beta_i\ , \gamma_i)$ has the form: It is required to find

$$\min g(\boldsymbol{r}) = \min \sum_i m_i(x_i^{\ \mathbf{2}} + y_i^{\mathbf{2}} + z_i^{\mathbf{2}}) \tag{3.3}$$

where the minimum is calculated over all states $(\boldsymbol{r}, \boldsymbol{v}) \in \Omega$ under the conditions:

$$L_x\ (\boldsymbol{r}, \boldsymbol{v}) = \sum_i m_i(\ y_i\gamma_i - z_i\beta_i) = 0 \tag{3.4}$$

$$L_y\ (\boldsymbol{r}, \boldsymbol{v}) = \sum_i m_i(\ z_i \propto_i - x_i\ \gamma_i) = 0 \tag{3.5}$$

$$L_z\ (\boldsymbol{r}, \boldsymbol{v}) = \sum_i m_i(\ x_i\beta_i - y_i \propto_i) = L \tag{3.6}$$

$$T(\boldsymbol{v}) = (1/2) \sum_i m_i(\propto_i^{\ \mathbf{2}} + \beta_i^{\ \mathbf{2}} + \gamma_i^{\ \mathbf{2}}) = T \tag{3.7}$$

where the parameters of this problem $L > 0$, $T > 0$. Next, we prove that the solution to the problem (3.3) – (3.7) and minimum size of the system (with non-zero angular momentum) at the current level of its kinetic energy can only be achieved on flat (but, in general, not necessarily finite) trajectories of the system, which partially coincides with some results obtained earlier in [9], but below we give a more precise proof of the transition from the above spatial problem (3.3) – (3.7) to the corresponding problem of nonlinear programming on the plane. And then, we will find the basic conditions under which the minimum size of the system at the current level of its kinetic energy is achieved on flat finite (circular or elliptical) trajectories, on which the flat structure of the system is preserved in time under the certain restrictions on the values of the system's integrals of motion. Let's denote by $\mathfrak{A}(L, T) \subset \Omega$ the subset of states $(\boldsymbol{r}, \boldsymbol{v}) \in \Omega$ of the phase space for which conditions (3.4) – (3.7) are satisfied.

**Theorem 3.1** *Let the minimum of the function $g(\boldsymbol{r})$ on the set $\mathfrak{A}(L, T) \subset \Omega$ is achieved at some point $(\boldsymbol{r}, \boldsymbol{v}) \in \mathfrak{A}(L, T)$, where the parameters $L > 0$, $T > 0$. Then at this point the equalities $z_i = 0$, $\gamma_i = 0$ are satisfied for each $i = 1, \ldots, N$.*

*Proof.* Let $\mathfrak{A}^*(L, T) \subset \Omega^* = E_{6N}$ be the closure of the set $\mathfrak{A}(L, T) \subset \Omega$. Consider first the problem of finding the minimum of the function $g(\boldsymbol{r})$ on the set $\mathfrak{A}^*(L, T)$. It is easy to show that this minimum is reached at some (perhaps not the only) point $(\boldsymbol{r}, \boldsymbol{v}) \in \mathfrak{A}^*(L, T)$. Indeed, let $\mathfrak{A}_C^*(L, T) \subset \mathfrak{A}^*(L, T)$ be a subset of all points that belong to $\mathfrak{A}^*(L, T)$ and such that $g(\boldsymbol{r}) \leq C$, where $C$ is any sufficiently large constant such that $C > inf\ g(\boldsymbol{r})$ on the set $\mathfrak{A}^*(L, T)$. The set $\mathfrak{A}_C^*(L, T)$ is closed due to the continuity of the functions $L_x\ (\boldsymbol{r}, \boldsymbol{v}), L_y\ (\boldsymbol{r}, \boldsymbol{v}), L_z\ (\boldsymbol{r}, \boldsymbol{v}), g(\boldsymbol{r}), T(\boldsymbol{v})$ and is bounded, taking into account also the condition (3.7) for the function $T(\boldsymbol{v})$. Thus, the function $g(\boldsymbol{r})$ is continuous on a closed bounded set $\mathfrak{A}_C^*(L, T)$, whence it follows that the minimum of the function $g(\boldsymbol{r})$ is reached on the set $\mathfrak{A}_C^*(L, T)$ and, consequently, this minimum is also reached on the set $\mathfrak{A}^*(L, T)$.

All functions in (3.3) – (3.7) are continuous and have continuous partial derivatives at each point $(\boldsymbol{r}, \boldsymbol{v}) \in \Omega^* = E_{6N}$. The necessary conditions for the minimum of the function $g = g(\boldsymbol{r})$ on the set $\mathfrak{A}^*(L, T) \subset \Omega^*$ at the point $(\boldsymbol{r}, \boldsymbol{v}) \in \mathfrak{A}^*(L, T)$ have the form:

$$\nabla g = a_1 \nabla L_x + a_2 \nabla L_y + a_3 \nabla L_z + \lambda \nabla T \tag{3.8}$$

where $a_1$, $a_2$, $a_3$, $\lambda$ are some coefficients (Lagrange multipliers). Note that the gradient vectors on the right side of (3.8) are linearly independent at any point $(\boldsymbol{r}, \boldsymbol{v}) \in \mathfrak{A}^*(L, T)$ due to the conditions of the problem $L > 0$, $T > 0$ (see below Remark 3.1), which ensures the correctness of the necessary minimum condition in (3.8). Taking into account that $\left(\partial T/_{\partial \boldsymbol{r}_i}\right) = \mathbf{0}, \left(\partial g/_{\partial \boldsymbol{v}_i}\right) = \mathbf{0},\ i = 1,\dots,N$; equations (3.8) are further written as

$$\frac{\partial g}{\partial \boldsymbol{r}_i} = a_1 \frac{\partial L_x}{\partial \boldsymbol{r}_i} + a_2 \frac{\partial L_y}{\partial \boldsymbol{r}_i} + a_3 \frac{\partial L_z}{\partial \boldsymbol{r}_i}, \quad i = 1,\dots,N \tag{3.9}$$

$$-\lambda \frac{\partial T}{\partial \boldsymbol{v}_i} = a_1 \frac{\partial L_x}{\partial \boldsymbol{v}_i} + a_2 \frac{\partial L_y}{\partial \boldsymbol{v}_i} + a_3 \frac{\partial L_z}{\partial \boldsymbol{v}_i}, \quad i = 1,\dots,N \tag{3.10}$$

Equations (3.9) in variables $(x_i, y_i, z_i)$, $(\propto_i, \beta_i, \gamma_i)$ have the form:

$$2x_i = -a_2\gamma_i + a_3\beta_i, \quad i = 1,\dots,N \tag{3.11}$$

$$2y_i = a_1\gamma_i - a_3 \propto_i, \quad i = 1,\dots,N \tag{3.12}$$

$$2z_i = -a_1\beta_i + a_2 \propto_i, \quad i = 1,\dots,N \tag{3.13}$$

Equations (3.10) in variables $(x_i, y_i, z_i)$, $(\propto_i, \beta_i, \gamma_i)$, are written as

$$-\lambda \propto_i = a_2 z_i - a_3 y_i, \quad i = 1,\dots,N \tag{3.14}$$

$$-\lambda\beta_i = -a_1 z_i + a_3 x_i, \quad i = 1,\dots,N \tag{3.15}$$

$$-\lambda\gamma_i = a_1 y_i - a_2 x_i, \quad i = 1,\dots,N \tag{3.16}$$

Let's introduce the vector of coefficients $\mathbf{a} = (a_1, a_2, a_3)$. Then these equations can be written in a more compact vector form:

$$2\boldsymbol{r}_i = [\boldsymbol{v}_i\mathbf{a}], \quad \lambda\boldsymbol{v}_i = [\boldsymbol{r}_i\mathbf{a}], \quad i = 1,\dots,N \tag{3.17}$$

where [ · ] is the vector multiplication. Or in matrix form:

$$2\boldsymbol{r}_i = \|A\|\, \boldsymbol{v}_i, \quad \lambda\boldsymbol{v}_i = \|A\|\, \boldsymbol{r}_i, \quad i = 1,\dots,N \tag{3.18}$$

where the matrix $\|A\|$ has the form:

$$\|A\| = \begin{Vmatrix} 0 & a_3 & -a_2 \\ -a_3 & 0 & a_1 \\ a_2 & -a_1 & 0 \end{Vmatrix}$$

From equalities (3.17) it follows that at any minimum point $(\boldsymbol{r}, \boldsymbol{v}, a_1, a_2, a_3, \lambda)$, $(\boldsymbol{r}, \boldsymbol{v}) \in \mathfrak{A}^*(L, T)$ the vector $\mathbf{a} \neq 0$. Indeed, if $\mathbf{a} = 0$, then by virtue of the first equality in (3.17) at the minimum point the equalities $\boldsymbol{r}_i = \mathbf{0}$ are satisfied for each $i = 1,\dots,N$; which contradicts the condition $L > 0$.

It also follows from (3.17) that at any point of the minimum $(\boldsymbol{r}, \boldsymbol{v}, a_1, a_2, a_3, \lambda)$, $(\boldsymbol{r}, \boldsymbol{v}) \in \mathfrak{A}^*(L, T)$ the coefficient $\lambda \neq 0$. Indeed, let $\lambda = 0$. Then at the minimum point any non-zero vector $\boldsymbol{r}_i$ (at least one such vector exists due to condition $L > 0$) is collinear to the vector **a** due to the second equality in (3.17). But by virtue of the first equality in (3.17), this same vector is perpendicular to the vector **a**, which means a contradiction. Thus, at any minimum point the coefficient $\lambda \neq 0$.

From (3.17) it also follows that at any minimum point $(\boldsymbol{r}, \boldsymbol{v}) \in \mathfrak{A}^*(L, T)$ for each particle of the system vectors $\boldsymbol{r}_i, \boldsymbol{v}_i$ are either simultaneously equal to zero or simultaneously not equal to zero. Note that if at the minimum point the state $(\boldsymbol{r}, \boldsymbol{v})$ of the system belongs to the phase space $\Omega$, that is $(\boldsymbol{r}, \boldsymbol{v}) \in \mathfrak{A}(L, T) \subset \Omega$, then equalities $\boldsymbol{r}_i = 0$, $\boldsymbol{v}_i = 0$ can be satisfied only for one (central) particle. From equations (3.17) it also follows that at any minimum point $(\boldsymbol{r}, \boldsymbol{v}) \in \mathfrak{A}^*(L,T)$ for each particle of the system with a non-zero radius vector $\boldsymbol{r}_i$, its velocity vector $\boldsymbol{v}_i$ is perpendicular to the radius vector.

Let's denote by $\mathbf{p_1} = (0, -a_3, a_2)$, $\mathbf{p_2} = (a_3, 0, -a_1)$, $\mathbf{p_3} = (-a_2, a_1, 0)$ the column vectors of the matrix $\|A\|$. Then, by virtue of (3.18), at any point of the minimum the following equalities hold:

$$2\boldsymbol{r}_i = \propto_i \mathbf{p_1} + \beta_i \mathbf{p_2} + \gamma_i \mathbf{p_3}\,, \quad i = 1, \dots, N \tag{3.19}$$

$$\lambda \boldsymbol{v}_i = x_i \mathbf{p_1} + y_i \mathbf{p_2} + z_i \mathbf{p_3}\,, \quad i = 1, \dots, N \tag{3.20}$$

The determinant of the matrix $\|A\|$ is equal to zero, and the vectors $\mathbf{p_1}$, $\mathbf{p_2}$, $\mathbf{p_3}$ are coplanar and these vectors, evidently, cannot be collinear. Thus, from (3.19), (3.20) it follows that at any point of the minimum $(\boldsymbol{r}, \boldsymbol{v}) \in \mathfrak{A}^*(L,T)$ all non-zero vectors $\boldsymbol{r}_i$, $\boldsymbol{v}_i$ are in the same plane spanned by coplanar vectors $\mathbf{p_1}$, $\mathbf{p_2}$, $\mathbf{p_3}$. And due to the definition of the angular momentum of the system, this plane must be perpendicular to the angular momentum vector $\boldsymbol{L} = (0, 0, L)$, i.e. the following equations are fulfilled: $(\mathbf{p_1}, \boldsymbol{L}) = a_2 L = 0$, $(\mathbf{p_2}, \boldsymbol{L}) = -a_1 L = 0$, $(\mathbf{p_3}, \boldsymbol{L}) = 0$. From these relations it follows that the coefficients $a_1, a_2$ satisfy the equalities: $a_1 = 0$, $a_2 = 0$, i.e. the above spatial problem of nonlinear programming in (3.3) – (3.7) is transformed into a plane problem of nonlinear programming (on a plane perpendicular to the angular momentum vector $\boldsymbol{L}$ of the system). And since the vector $\boldsymbol{L}$ is directed along the axis z, this plane is the plane $z = 0$.

Thus, at any point of the minimum $(\boldsymbol{r}, \boldsymbol{v}, a_1, a_2, a_3, \lambda), (\boldsymbol{r}, \boldsymbol{v}) \in \mathfrak{A}^*(L,T)$ all vectors $\boldsymbol{r}_i$, $\boldsymbol{v}_i$, $i = 1, \dots, N$ are in the same plane z $= 0$ and coefficients $a_1 = a_2 = 0$. Then the necessary conditions of the minimum in (3.11) – (3. 16) for the coordinates of vectors $\boldsymbol{r}_i = (x_i, y_i)$, $\boldsymbol{v}_i = (\propto_i, \beta_i)$ on this plane take the following form:

$$2x_i = a_3 \beta_i\,, \quad 2y_i = -a_3 \propto_i\,, \quad i = 1, \dots, N \tag{3.21}$$

$$\lambda \propto_i = a_3 y_i\,, \quad \lambda \beta_i = -a_3 x_i\,, \quad i = 1, \dots, N \tag{3.22}$$

After which we find that the necessary conditions for the minimum in this problem have the following form: $a_1 = a_2 = 0$, $2\lambda = -(a_3)^2 \neq 0$, and

$$z_i = 0, \quad \gamma_i = 0, \quad i = 1, \dots, N \tag{3.23}$$

$$\propto_i = -(2/a_3) y_i\,, \quad \beta_i = (2/a_3) x_i\,, \quad i = 1, \dots, N \tag{3.24}$$

Conditions (3.6), (3.7) are then written in the form of the following equalities:

$$\textstyle\sum_i m_i (x_i \beta_i - y_i \propto_i) = (2/a_3) \sum_i m_i (x_i^2 + y_i^2) = L$$

$$\textstyle\sum_i m_i (x_i \beta_i - y_i \propto_i) = (a_3/2) \sum_i m_i (\propto_i^2 + \beta_i^2) = L$$

$$\textstyle(1/2) \sum_i m_i (\propto_i^2 + \beta_i^2) = (2/a_3^2) \sum_i m_i (x_i^2 + y_i^2) = T$$

from which follows the equality $a_3 = L/T$ and we finally find that in problem (3.3) – (3.7) the necessary conditions for the minimum at point $(\boldsymbol{r}, \boldsymbol{v}) \in \mathfrak{A}^*(L,T)$ have the form:

$$z_i = 0, \quad \gamma_i = 0, \quad i = 1, \dots, N \tag{3.25}$$

$$\propto_i = -(2T/L) y_i\,, \quad \beta_i = (2T/L) x_i\,, \quad i = 1, \dots, N \tag{3.26}$$

$$\textstyle\sum_i m_i (x_i^2 + y_i^2) = L^2/(2T) \tag{3.27}$$

Whence also it follows that the minimum of the function $g(\boldsymbol{r})$ on the set $\mathfrak{A}^*(L,T)$ is achieved at any point that satisfies the conditions (3.25) – (3.27), and this minimum is equal:

$$\min g(\boldsymbol{r}) = L^2/(2T) \tag{3.28}$$

For some points $(\boldsymbol{r},\boldsymbol{v}) \in \mathfrak{A}^*(L,T)$ that satisfy the necessary minimum conditions (3.25) – (3.27), the radius-vectors $\boldsymbol{r}_i = (x_i\,, y_i)$ of different particles can coincide. Such points belong to the closure $\mathfrak{A}^*(L,T)$, but do not belong to the subset $\mathfrak{A}(L,T) \subset \Omega$, over which the minimum is taken in problem (3.3) – (3.7). But as can be seen from these relations, conditions (3.25) – (3.27) are also satisfied at some points $(\boldsymbol{r},\boldsymbol{v}) \in \mathfrak{A}(L,T) \subset \Omega$ (i.e. at points with different vectors $\boldsymbol{r}_i$). For this, taking into account (3.26), it is sufficient of the existence of non-coinciding vectors $\boldsymbol{r}_i = (x_i\,, y_i)$ satisfying equality (3.27). Therefore, the minimum (3.28) is also achieved on the set $\mathfrak{A}(L,T) \subset \Omega$, from which, taking into account the necessary conditions for the minimum in (3.25), Theorem 3.1 follows. ■

**Remark 3.1** The gradient vectors on the right side of (3.8) are linearly independent at any point $(\boldsymbol{r},\boldsymbol{v}) \in \mathfrak{A}^*(L,T)$ due to the conditions of the problem $L > 0$, $T > 0$. Indeed, if these vectors are linearly dependent at some point $(\boldsymbol{r},\boldsymbol{v}) \in \mathfrak{A}^*(L,T)$, then at this point the right-hand side of (3.8) is equal to zero for some $\mathbf{a} = (a_1, a_2, a_3)$, $\lambda$. From this, similar to (3.9) – (3.17), the equations follow: $[\boldsymbol{v}_i\mathbf{a}] = \mathbf{0}$, $\lambda\boldsymbol{v}_i = [\boldsymbol{r}_i\mathbf{a}]$, $i = 1,\ldots,N$. But these equations have a single nontrivial solution: $\mathbf{a} \neq \mathbf{0}$ , $\lambda = 0$ and then all non-zero vectors $\boldsymbol{r}_i$, $\boldsymbol{v}_i$ are collinear to vector $\mathbf{a}$, which contradicts condition $L > 0$. And trivial solutions: 1) if $\lambda \neq 0$, $\mathbf{a} = \mathbf{0}$, then by virtue of the second equation all vectors $\boldsymbol{v}_i = \mathbf{0}$, $i = 1,\ldots,N$; which contradicts conditions $L > 0$, $T > 0$. 2) If $\lambda \neq 0$, $\mathbf{a} \neq \mathbf{0}$, then by virtue of the first equation any non-zero vector $\boldsymbol{v}_i$ is collinear to vector $\mathbf{a}$, and by virtue of the second equation this same vector $\boldsymbol{v}_i$ is perpendicular to vector $\mathbf{a}$, which means a contradiction, that is in this case also all vectors $\boldsymbol{v}_i = \mathbf{0}$, $i = 1,\ldots,N$; which also contradicts the conditions $L > 0, T > 0$. Thus, at any point $(\boldsymbol{r},\boldsymbol{v}) \in \mathfrak{A}^*(L,T)$, gradient vectors on the right side of (3.8) are linearly independent due to the conditions $L > 0$, $T > 0$.

**Remark 3.2** The functions on the right-hand side of the above system of differential equations (1.1), (1.2) are continuous and have continuous partial derivatives at each point $(\boldsymbol{r},\boldsymbol{v}) \in \Omega \subset E_{6N}$, i.e. these functions satisfy the known conditions ([8] et. al.) of existence and uniqueness of a solution (trajectory), starting at some moment of time $t_0$ from the initial state $(\boldsymbol{r},\boldsymbol{v}) \in \Omega$, on some time interval $(\tau_1,\ \tau_2)$ such that $-\infty \le \tau_1 < t_0 < \tau_2 \le \infty$. From this, taking into account the above Theorem 3.1, it follows that the trajectory, starting at moment of time $t_0$ from point $(\boldsymbol{r}^*,\boldsymbol{v}^*) \in \mathfrak{A}(L,T)$ of the minimum of the function $g(\boldsymbol{r})$ on the set $\mathfrak{A}(L,T) \subset \Omega$, is flat (in the plane $z = 0$), but this trajectory, in general is not necessarily finite, depending on the initial point $(\boldsymbol{r}^*,\boldsymbol{v}^*)$.

Thus, from Theorem 3.1 it follows that the minimum possible size of a system (with non-zero angular momentum $L > 0$) at the current value of its kinetic energy $T > 0$ can be achieved only on flat, not necessarily finite, trajectories. Next, we will find the basic conditions under which these trajectories are flat finite (elliptical) trajectories on which the flat structure of the system is preserved in time (up to rotation and scale).

Let $\boldsymbol{r} = (\boldsymbol{r}_1,\ \boldsymbol{r}_2, \ldots,\ \boldsymbol{r}_N) \in \mathfrak{R}_2$ be the complete set of radius vectors of all particles of the system on the plane, where $\boldsymbol{r}_i = (x_i\,, y_i)$ is the radius vector of the particle $i$ and $\mathfrak{R}_2 = \{\boldsymbol{r} :\ |\boldsymbol{r}_j - \boldsymbol{r}_i| > 0\,,\ \ i \neq j\ \} \subset E_{2N}$ is the coordinate space of the system on the plane. Let's denote by $\mathfrak{S}(g) = \{\boldsymbol{r} \in \mathfrak{R}_2 : g(\boldsymbol{r}) = g\} \subset \mathfrak{R}_2$ the subset of the system states on this plane with a fixed value of the function $g(\boldsymbol{r}) = g$. And let $\mathfrak{A}_\mathrm{m}(L,T) \subset \mathfrak{A}(L,T) \subset \Omega$ be the subset of states $(\boldsymbol{r},\boldsymbol{v}) \in \Omega$ of the phase space for which the necessary conditions (3.25) – (3.27) for the minimum in the above spatial problem (3.3) – (3.7) are satisfied. In accordance with Theorem 3.1, the minimum of the function $g(\boldsymbol{r})$ on the set $\mathfrak{A}(L,T) \subset \Omega$ is equal to the value $g = L^2/(2T)$ and is achieved at each point $(\boldsymbol{r}^*,\boldsymbol{v}^*) \in \mathfrak{A}_\mathrm{m}(L,T)$. And by virtue of conditions (3.26) at each minimum point $(\boldsymbol{r}^*, \boldsymbol{v}^*) \in \mathfrak{A}_{\mathrm{m}}(L,T)$ the vector $\boldsymbol{v}^* = (\boldsymbol{v}_1^*, \; . \; . \; . \; , \boldsymbol{v}_N^*)$ of particle velocities is uniquely determined by the vector of their coordinates $\boldsymbol{r}^* = (\boldsymbol{r}_1^*, \; . \; . \; . \; , \boldsymbol{r}_N^* \;) \in \mathfrak{S}(g) \subset \mathfrak{R}_2$ on the plane, where $g = L^2/(2T)$. Due to the necessary minimum conditions (3.25) – (3.27) the trajectory of the system starting from any point of the minimum $(\boldsymbol{r}^*, \boldsymbol{v}^*) \in \mathfrak{A}_{\mathrm{m}}(L,T)$ is flat, but this trajectory is not necessarily finite depending on the initial point $\boldsymbol{r}^* \in \mathfrak{S}(g) \subset \mathfrak{R}_2$, where $g = L^2/(2T)$.

Next, we define the initial point $\boldsymbol{r}^*$ on the set $\mathfrak{S}(g)$, $g = L^2/(2T)$, so that at this point the minimum of the potential function $f(\boldsymbol{r})$ is achieved on this set, i.e. from the condition:

$$f(\boldsymbol{r}^*) = \min f(\boldsymbol{r}), \quad \boldsymbol{r} \in \mathfrak{S}(g), \;\; g = L^2/(2T) \tag{3.29}$$

where the minimum is calculated over all states $\boldsymbol{r} \in \mathfrak{S}(g) \subset \mathfrak{R}_2$, $g = L^2/(2T)$. Let's introduce also the function:

$$h(g) \; = \min f(\boldsymbol{r}), \quad \boldsymbol{r} \in \mathfrak{S}(g), \tag{3.30}$$

where the minimum is calculated over all states $\boldsymbol{r} \in \mathfrak{S}(g) \subset \mathfrak{R}_2$. In accordance with [7], par.4 the solution $h(g)$ of this problem of nonlinear programming depending on the parameter of this problem $g > 0$ has the form (for any $N > 1$, $m_i > 0$, $i = 1, \dots, N$):

$$h(g) = C(\boldsymbol{m})/\sqrt{g} \qquad \text{for each} \;\; g > 0 \tag{3.31}$$

where $C(\boldsymbol{m}) > 0$ is a function of the masses of particles of the system, for which the following lower estimate ([7], par.5) holds:

$$C(\boldsymbol{m}) \geq \gamma \left(\sum_{i<j} m_i m_j\right)^{3/2} / \left(\sum_j m_j\right)^{1/2} \tag{3.32}$$

And the right-hand side of this inequality gives the exact value of the function $C(\boldsymbol{m})$ for cases $N = 2$ and $N = 3$ ([7], par.6). It's not difficult to show that the global minimum in the problem (3.30) is achieved for any $N > 1$, $m_i > 0$, $i = 1, \dots, N$ at some, not necessarily unique, point $\boldsymbol{r}^* = (\boldsymbol{r}_1^*, \; . \; . \; . \; , \boldsymbol{r}_N^*) \in \mathfrak{S}(g) \subset \mathfrak{R}_2$, at which the following necessary conditions for a minimum hold:

$$\boldsymbol{F}_i(\boldsymbol{r}^*) = -2\lambda m_i \boldsymbol{r}_i^*, \quad i = 1, \dots, N \tag{3.33}$$

where the coefficient (Lagrange multiplier) $\lambda > 0$ is related to the parameter $g > 0$ of this problem by the relation: $g = [C(\boldsymbol{m})/2\lambda]^{2/3}$. And at the point $\boldsymbol{r}^* \in \mathfrak{S}(g) \subset \mathfrak{R}_2$ of global minimum the characteristics $g(\boldsymbol{r}^*)$, $f(\boldsymbol{r}^*)$ satisfy the following relations ([7], par.4, par.7):

$$g(\boldsymbol{r}^*) = [C(\boldsymbol{m})/2\lambda]^{2/3}, \quad f(\boldsymbol{r}^*) = C(\boldsymbol{m})/\sqrt{g(\boldsymbol{r}^*)} \; = [C(\boldsymbol{m})]^{2/3}(2\lambda)^{1/3} \tag{3.34}$$

Next, determine the vector $\boldsymbol{v}^* = (\boldsymbol{v}_1^*, \; . \; . \; . \; , \boldsymbol{v}_N^* \;)$ of particle velocities at the initial point $(\boldsymbol{r}^*, \boldsymbol{v}^*)$ of the trajectory, taking into account (3.26), as follows: $\boldsymbol{v}_i^* = (\propto_i^*, \beta_i^*)$, where

$$\propto_i^* = -(2T/L) y_i^*, \quad \beta_i^* = (2T/L)\, x_i^*, \quad i = 1, \dots, N \tag{3.35}$$

where $x_i^*, y_i^*$ are the coordinates of the radius vector $\boldsymbol{r}_i^* = (x_i^*, y_i^*)$, $i = 1, \dots, N$. Then point $(\boldsymbol{r}^*, \boldsymbol{v}^*)$ belongs to subset $\mathfrak{A}_{\mathrm{m}}(L,T) \subset \mathfrak{A}(L,T) \subset \Omega$. And by virtue of (3.33), (3.35), the following relations are valid at this point:

$$\boldsymbol{F}_i(\boldsymbol{r}^*) = -2\lambda m_i \boldsymbol{r}_i^*, \quad i = 1, \dots, N \tag{3.36}$$

$$\propto_i^* = -\omega_0 y_i^*, \quad \beta_i^* = \omega_0\, x_i^*, \quad i = 1, \dots, N \tag{3.37}$$

where $\omega_0 = 2T/L > 0$, $\lambda = C(\boldsymbol{m})/(2g^{3/2}) > 0$, $g = L^2/(2T)$. It also follows that at this point the equalities: $\sum_i m_i \boldsymbol{r}_i^* = \boldsymbol{0}$, $\sum_i m_i \boldsymbol{v}_i^* = \boldsymbol{0}$ are automatically satisfied.

Relations (3.36), (3.37) define a flat trajectory of the form $\tilde{\boldsymbol{r}}_\lambda(t, \omega_0),\ t \geq 0$ ([7], par.8) with parameters $\lambda$, $\omega_0$ and initial point $(\boldsymbol{r}^*, \boldsymbol{v}^*) \in \mathfrak{A}_{\mathrm{m}}(L, T)$. If the trajectory parameter $P = \omega_0^2/(2\lambda)$ satisfies the inequalities $0 < P < 2$, then this trajectory is finite and elliptical with eccentricity $e = 1 - P$, on which the flat structure of the system is preserved in time (up to rotation and scale). From this, we find that if the parameters $L > 0, T > 0$ satisfy inequality:

$$P = \omega_0^2/(2\lambda) = L\sqrt{2T}/C(\boldsymbol{m}) < 2 \qquad (3.38)$$

then the minimum possible size of the system (with a non-zero angular momentum $L > 0$) at the current value of its kinetic energy $T > 0$ is achieved at the initial point $(\boldsymbol{r}^*, \boldsymbol{v}^*) \in \mathfrak{A}_{\mathrm{m}}(L, T)$ of the above flat finite elliptic trajectory, on which the minimum of the function $f(\boldsymbol{r})$ is achieved on the set $\mathfrak{S}(g) \subset \mathfrak{R}_2,\ g = L^2/(2T)$.

From relations (3.34), (3.37) after simple transformations it follows that the characteristics of a system at the initial point $(\boldsymbol{r}^*, \boldsymbol{v}^*) \in \mathfrak{A}_{\mathrm{m}}(L, T)$ of this trajectory are expressed through the parameter $P = \omega_0^2/(2\lambda)$ of this trajectory as follows:

$$f(\boldsymbol{r}^*) = [C(\boldsymbol{m})]^{2/3} {\omega_0}^{2/3} / P^{1/3} \qquad (3.39)$$

$$T(\boldsymbol{v}^*) = \sum_i m_i (\boldsymbol{v}_i^*)^2/2 = \omega_0^2\, g(\boldsymbol{r}^*)/2 = f(\boldsymbol{r}^*)P/2 \qquad (3.40)$$

$$L(\boldsymbol{r}^*, \boldsymbol{v}^*) = \sum_i m_i |\boldsymbol{v}_i^*||\boldsymbol{r}_i^*| = \omega_0\, g(\boldsymbol{r}^*) = [C(\boldsymbol{m})]^{2/3} P^{2/3} / {\omega_0}^{1/3} \qquad (3.41)$$

And, due to the relations (3.28), (3.31), (3.37), at this point the equalities are satisfied:

$$T(\boldsymbol{v}^*) = (1/2) \sum_i m_i [(\alpha_i^*)^2 + (\beta_i^*)^2] = T,\ f(\boldsymbol{r}^*) = C(\boldsymbol{m})/\sqrt{g} = C(\boldsymbol{m})\sqrt{2T}/L \qquad (3.42)$$

It follows that inequality in (3.38) is also equivalent to the following inequality:

$$T(\boldsymbol{v}^*) < f(\boldsymbol{r}^*), \qquad (3.43)$$

where $T(\boldsymbol{v}^*),\ f(\boldsymbol{r}^*)$ are the values of the system kinetic and potential energy functions at this point. Taking into account the lower bound for function $C(\boldsymbol{m})$ in (3.32), inequality in (3.38) or (3.43) is valid if the following simpler condition is satisfied:

$$L\sqrt{T} < \sqrt{2}\,\gamma \left(\sum_{i<j} m_i m_j\right)^{3/2} / \left(\sum_j m_j\right)^{1/2} \qquad (3.44)$$

From (3.42), (3.43) it also follows that on this trajectory the system total energy $E = T - f$ satisfies the relations:

$$T(\boldsymbol{v}^*) - f(\boldsymbol{r}^*) = T - \frac{C(\boldsymbol{m})\sqrt{2T}}{L} = E < 0, \qquad (3.45)$$

from which it follows, in particular, that at fixed values of the integrals of motion: $L > 0$, $E < 0$ the minimum possible size of the system at the current value of its kinetic energy $T$ can only be achieved for values of $T > 0$ that satisfy equation (3.45) (see also formulas (3.51) and Remark 3.3 below).

Note that the minimum of the left-hand side of equation (3.45) for $T > 0$ at a fixed value of $L > 0$ is equal to the value: $E_{min} = -[C(\boldsymbol{m})]^2/(2L^2) < 0$, which corresponds to formula (2.23) in the previous item 2. From this it follows, by virtue of equation (3.45), that the minimum possible size of the system at the current value of its kinetic energy $T > 0$ can be achieved only if the total energy $E$ of the system is in the interval: $-[C(\boldsymbol{m})]^2/(2L^2) \leq E < 0$, which coincides with the following main condition (3.47) for the integrals of motion of the system $L > 0, E < 0$.

The equation in (3.45) after simple transformations can be written, taking into account (3.38), in the following form:

$$P(2-P) = -\frac{2EL^2}{[C(\boldsymbol{m})]^2} \tag{3.46}$$

where $P = \omega_0^2/(2\lambda) = L\sqrt{2T}/C(\boldsymbol{m})$ is the parameter of the above flat elliptical trajectory. Let's introduce parameter $W = W(L,E) = -2EL^2/[C(\boldsymbol{m})]^2$. If the characteristics $L > 0$, $E < 0$ satisfy the inequalities:

$$0 < -2EL^2 \leq [C(\boldsymbol{m})]^2, \tag{3.47}$$

then equation (3.46) has two solutions:

$$P_1 = 1 - \sqrt{1 - W(L,E)}, \quad P_2 = 1 + \sqrt{1 - W(L,E)} \tag{3.48}$$

where $0 < P_1 < P_2 < 2$. Thus, if the system characteristics (integrals of motion) $L > 0$, $E < 0$ satisfy the condition (3.47), then the minimum possible size of the system at the current value of its kinetic energy $T > 0$ is achieved at the initial points of two above flat finite elliptical trajectories of the form $\tilde{\boldsymbol{r}}_\lambda(t,\omega_0)$ with parameters $\lambda_1$, $\omega_{01}$, $P_1 = \omega_{01}^2/(2\lambda_1)$ and $\lambda_2$, $\omega_{02}$, $P_2 = \omega_{02}^2/(2\lambda_2)$ with symmetrical eccentricity parameters: $e = \pm\sqrt{1 - W(L,E)}$.

Next, from relations (3.34), (3.37), we find the parameters $\lambda, \omega_0$ of these two trajectories depending on the values of $L > 0$, $E < 0$:

$$\lambda_1 = P_1^3[C(\boldsymbol{m})]^4/(2L^6), \quad \omega_{01} = P_1^2[C(\boldsymbol{m})]^2/L^3 \tag{3.49}$$

$$\lambda_2 = P_2^3[C(\boldsymbol{m})]^4/(2L^6), \quad \omega_{02} = P_2^2[C(\boldsymbol{m})]^2/L^3 \tag{3.50}$$

where the parameters $P_1$, $P_2$ are defined above in (3.48). Taking into account the relations (3.39) – (3.41), we also find the values of the kinetic and potential energy functions $T, f$ at the initial points of these two trajectories, depending on the characteristics $L > 0$, $E < 0$:

$$T_1 = P_1^2[C(\boldsymbol{m})]^2/(2L^2) = |E|\, P_1/P_2, \quad T_2 = P_2^2\,[C(\boldsymbol{m})]^2/(2L^2) = |E|\, P_2/P_1 \tag{3.51}$$

$$f_1 = P_1\,[C(\boldsymbol{m})]^2/L^2 = 2|E|/P_2, \quad f_2 = P_2\,[C(\boldsymbol{m})]^2/L^2 = 2|E|/P_1 \tag{3.52}$$

**Remark 3.3** Next, taking into account the relations (3.34), we also find the values of the function $g = g(\boldsymbol{r})$ at the initial points of these two trajectories depending on the characteristics $L > 0$, $E < 0$:

$$g_1 = P_1^{-2}L^4/[C(\boldsymbol{m})]^2, \quad g_2 = P_2^{-2}L^4/[C(\boldsymbol{m})]^2 \tag{3.53}$$

From (3.51) – (3.53) the equalities also follow: $g_1 = L^2/(2T_1)$, $g_2 = L^2/(2T_2)$, which correspond to the solution of the above spatial problem in Theorem 3.1 of finding the minimum of the function $g(\boldsymbol{r})$ at fixed values of the angular momentum $L = |\boldsymbol{L}| > 0$ and kinetic energy $T > 0$. That is, at these points the minimum possible size of the system at the current value of its kinetic energy is achieved.

**Remark 3.4** The lower estimate for the function $C(\boldsymbol{m})$ on the right side in (3.32) allows us to find the following simpler sufficient condition for the fulfillment of the main condition (3.47) for a system with particle masses $\boldsymbol{m} = (m_1, m_2, \dots, m_N)$:

$$0 < -2EL^2 \leq \gamma^2\left(\sum_{i<j} m_i m_j\right)^3 / \sum_j m_j \tag{3.54}$$

Thus, if the system characteristics (integrals of motion) $L > 0$, $E < 0$ satisfy the condition (3.47) or the simpler condition (3.54), then the minimum possible size of the system at the current value of its kinetic energy is achieved at the initial points of two above flat finite elliptical trajectories of the form $\tilde{\boldsymbol{r}}_\lambda(t, \omega_0)$ with parameters $\lambda_1$, $\omega_{01}$ , $P_1 = \omega_{01}^2/(2\lambda_1)$ and $\lambda_2$, $\omega_{02}$ , $P_2 = \omega_{02}^2/(2\lambda_2)$ with symmetrical eccentricity parameters:

$$e = \pm\sqrt{1 - W(L, E)} \tag{3.55}$$

**Remark 3.5** Due to symmetry relations (3.48), (3.55) these trajectories are actually two versions of the same flat elliptical trajectory, which start from two different (opposite) points on this trajectory. Namely, on the same (for example, first) flat elliptical trajectory of type $\tilde{\boldsymbol{r}}_\lambda(t, \omega_0)$ with parameters $\lambda_1$, $\omega_{01}$, $P_1 = \omega_{01}^2/(2\lambda_1) = 1 - \sqrt{1 - W(L, E)}$ and eccentricity:

$$e = \sqrt{1 - W(L, E)} = \sqrt{1 - \frac{2|E|L^2}{[C(\boldsymbol{m})]^2}} \tag{3.56}$$

And the minimum possible size of the system at the current value of its kinetic energy is achieved on this trajectory at the points with angular coordinates $\varphi = 2l\pi,\ l = 0,1,2, \ldots$ with characteristics $T_1, f_1, g_1$ and at the points with opposite angular coordinates $\varphi = (2l + 1)\pi,\ l = 0,1,2, \ldots$ with characteristics $T_2,\ f_2, g_2$.. If the parameter $W(L, E)$ =1, then the minimum possible size of the system is achieved at all points of this flat finite "circular" trajectory with parameters $P_1 = P_2 = 1$ and eccentricity $e = 0$. At this case the formulars for the system characteristics in (3.51) – (3.53) coincide with the formulars (2.24) obtained above in Section 2 for the spatial problem of finding local extrema of the system total energy function $E$.

Note that relations of the form (3.33) are fulfilled not only for the above point $\boldsymbol{r}^* = (\boldsymbol{r}_1^*, \ . \ . \ . \ , \boldsymbol{r}_N^*) \in \mathfrak{S}(g) \subset \mathfrak{R}_2$ of the global minimum in problem (3.30), but also for the point $\boldsymbol{r}_k^* = (\boldsymbol{r}_{k1}^*, \ . \ . \ . \ , \boldsymbol{r}_{kN}^*) \in \mathfrak{S}(g) \subset \mathfrak{R}_2$ of any kth local minimum in this problem:

$$\boldsymbol{F}_i(\boldsymbol{r}_k^*) = -2\lambda_k m_i \boldsymbol{r}_{ki}^*, \quad i = 1, \ldots, N; \quad k = 1, \ldots, n \tag{3.57}$$

It follows that the above relations and results are also valid for the case of any kth local minimum in this problem, up to replacing the function $C(\boldsymbol{m})$ with the function $C_k(\boldsymbol{m})$, for which the following lower estimates ([7], par.7) are satisfied:

$$C_k(\boldsymbol{m}) \geq C(\boldsymbol{m}) \geq \gamma\left(\textstyle\sum_{i<j} m_i m_j\right)^{3/2} / \left(\textstyle\sum_j m_j\right)^{1/2}, \ k = 1, \ldots, n \tag{3.58}$$

Repeating the previous relations, we find that if the system characteristics (integrals of motion) $L > 0$, $E < 0$ satisfy the condition:

$$0 < -2EL^2 \leq [C_k(\boldsymbol{m})]^2, \tag{3.59}$$

then the minimum possible size of the system at the current value of its kinetic energy is achieved on the flat finite elliptic trajectory of the form $\tilde{\boldsymbol{r}}_\lambda(t, \omega_0)$ with parameters:

$$\lambda_k = P_{1k}^3 [C_k(\boldsymbol{m})]^4/(2L^6), \quad \omega_{0k} = P_{1k}^2 [C_k(\boldsymbol{m})]^2/L^3 \tag{3.60}$$

with eccentricity $e_k = \sqrt{1 - W_k(L, E)}$, where $W_k(L, E) = -2EL^2/[C_k(\boldsymbol{m})]^2$,

$$P_{1k} = 1 - \sqrt{1 - W_k(L, E)}\,, \ P_{2k} = 1 + \sqrt{1 - W_k(L, E)}\,, \ k = 1, \ldots, n.$$

And the minimum possible size of the system at the current value of its kinetic energy is achieved on this trajectory at the points with angular coordinates $\varphi = 2l\pi$ and $\varphi = (2l+1)\pi,\ l = 0,1,2,\dots$ with characteristics $T, f, g$ of the following form:

$$T_{1k} = P_{1k}^2[C_k(\boldsymbol{m})]^2/(2L^2) = |E|\,P_{1k}/P_{2k} \tag{3.61}$$

$$f_{1k} = P_{1k}\,[C_k(\boldsymbol{m})]^2/L^2 = 2|E|/P_{2k} \tag{3.62}$$

$$T_{2k} = P_{2k}^2\,[C_k(\boldsymbol{m})]^2/(2L^2) = |E|\,P_{2k}/P_{1k} \tag{3.63}$$

$$f_{2k} = P_{2k}\,[C_k(\boldsymbol{m})]^2/L^2 = 2|E|/P_{1k} \tag{3.64}$$

$$g_{1k} = P_{1k}^{-2}L^4/[C_k(\boldsymbol{m})]^2, \quad g_{2k} = P_{2k}^{-2}L^4/[C_k(\boldsymbol{m})]^2 \tag{3.65}$$

where $k = 1,\dots,n$. If the parameter $W_k(L,E)$ =1, then the minimum possible size of the system at the current value of its kinetic energy is achieved at all points of this flat finite "circular" trajectory with parameters $P_{1k} = P_{2k} = 1$ and eccentricity $e_k = 0$. At this case the formulars for the system characteristics in (3.61) – (3.65) coincide with formulars (2.26), (2.27), obtained above in Section 2 for the spatial problem of finding local extrema of the system total energy function $E$.

**Remark 3.6** Since inequality $C_k(\boldsymbol{m}) \geq C(\boldsymbol{m})$ holds for any kth local minimum ([7], par.7), all inequalities in (3.59) for the case of local minima are satisfied if the basic condition (3.47) or the simpler sufficient condition (3.54) is satisfied.

**Remark 3.7** Similarly, almost repeating the previous relations, the similar results are proved, namely: the minimum values of the functions $T, f$ of the kinetic and potential energy of the system (with non-zero angular momentum) at the current value of its size can only be achieved on flat, in general, not necessarily finite, trajectories of the system's motion in space. And if the integrals of motion $L > 0, E < 0$ satisfy the condition (3.47) or the simpler condition (3.54), then the minimum values of the functions $T, f$ at the current value of the system size are achieved at the same "extreme points" with opposite angular coordinates $\varphi = 2l\pi$ and $\varphi = (2l+1)\pi,\ l = 0,1,2,\dots$ on the flat finite trajectories of the form $\tilde{\boldsymbol{r}}_\lambda(t,\omega_0)$ with parameters (3.60). And the system characteristics $T, f, g$ at these points take the values indicated above in (3.61) – (3.65).

Thus, if conditions (3.47) or (3.54), which limit the values of the integrals of motion $L > 0$, $E < 0$, are satisfied, then these minimum possible (in the above sense) values of the system characteristics T, f, $g$ are determined by the equalities in (3.61) – (3.65) and are achieved on the above flat finite (elliptic or circular) trajectories. And on these trajectories the flat structure of the system is preserved in time (up to rotation and scale) under the above restrictions on the values of the integrals of motion of the system $L > 0$, $E < 0$.

In the general case, the number of such flat finite trajectories, on which the minimum possible values of the system characteristics T, f, $g$ are achieved and the flat structure of the system is preserved in time, is equal to q = $q(L,E)$, where q$(L,E)$ is the number of indices $k$ for which the integrals of motion $L > 0, E < 0$ satisfy the conditions (3.59).

**Remark 3.8** The previous relations depend on the functions $C(\boldsymbol{m}), C_k(\boldsymbol{m})$ which in the general case at $N > 3$ are quite complex. However, using these functions and the lower bounds for these functions in (3.32), (3.58) allows us to find essential relations for the characteristics of the system, including the main conditions (3.47), (3.54), (3.59) on the values of the integrals of motion: $L > 0$ , $E < 0$ under which the above minimum possible values of the system characteristics T, f, $g$ are achieved on the above flat finite trajectories, (on which the flat structure of the system is preserved in time under the above conditions), as well as estimates ([7], par.5, etc.) for the system characteristics.

**Remark 3.9** Let $N = 3$. Then the flat finite trajectories, found above on the basis of the solution of the spatial problem of nonlinear programming in Theorem 3.1, coincide at this case with the well-known Lagrange solution for the three-body problem [2], [3], in which the system structure has the shape of an equilateral triangle (see also [7], par. 6).

## Appendix

Let us consider a more general model in which the interaction force $\boldsymbol{F}_{ij} = \boldsymbol{F}_{ij}(\boldsymbol{r})$ has the form:

$$\boldsymbol{F}_{ij}(\boldsymbol{r}) = \frac{\gamma_p\, m_i\, m_j\, \boldsymbol{L_{ij}}}{\left|\boldsymbol{L_{ij}}\right|^{1+p}}, \quad i \neq j, \quad (i,j) \in (1, \dots, N) \tag{A.1}$$

where $p > 1$, $\gamma_p > 0$, $\boldsymbol{r} = (\boldsymbol{r}_1, \boldsymbol{r}_2, \dots, \boldsymbol{r}_N) \in \Re$, $\boldsymbol{L_{ij}} = \boldsymbol{L_{ij}}(\boldsymbol{r}) = \boldsymbol{r}_j - \boldsymbol{r}_i$. The interaction force $\boldsymbol{F}_i(\boldsymbol{r})$ acting on the particle $i$ from all other particles of the system has the form:

$$\boldsymbol{F}_i(\boldsymbol{r}) = \sum_{j\neq i} \boldsymbol{F}_{ij}(\boldsymbol{r}) = \sum_{j\neq i} \frac{\gamma_p\, m_i\, m_j\, \boldsymbol{L_{ij}}}{\left|\boldsymbol{L_{ij}}\right|^{1+p}}, \quad \boldsymbol{r} \in \Re \tag{A.2}$$

If the parameter $p = 2$, then formulas (A.1), (A.2) correspond to Newton's law of gravity, when the force of attraction between two particles with masses $m_i, m_j$ is inversely proportional the square of the distance between them.

The trajectory of the system is determined by a system of differential equations similar to the above equations (1.1), (1.2):

$$m_i \dot{\boldsymbol{v}}_i = \boldsymbol{F}_i(\boldsymbol{r}), \quad i = 1, \dots, N \tag{A.3}$$

$$\dot{\boldsymbol{r}}_i = \boldsymbol{v}_i, \quad i = 1, \dots, N \tag{A.4}$$

with some initial conditions, where the functions $\boldsymbol{F}_i(\boldsymbol{r})$ have the form (A.2). The potential function $f(\boldsymbol{r})$ has the form:

$$f(\boldsymbol{r}) = \frac{1}{2(p-1)} \sum_i \sum_{j\neq i} \frac{\gamma_p\, m_i\, m_j}{\left|\boldsymbol{L_{ij}}\right|^{p-1}}, \quad \boldsymbol{r} \in \Re \tag{A.5}$$

Just as above in Section 1, a trajectory $[\boldsymbol{r}(t), \boldsymbol{v}(t)] \in \Omega$, $t \geq 0$ satisfying the system of differential equations (A.3), (A.4) under some initial conditions will be called finite if there exist constants $0 < C_1 < C_2 < \infty$ such that $C_1 < \left|\boldsymbol{r}_j(t) - \boldsymbol{r}_i(t)\right| < C_2$ for each $i \neq j$, $t \geq 0$. And this trajectory will be called regular if there exist constant $C > 0$ such that $\left|\boldsymbol{r}_j(t) - \boldsymbol{r}_i(t)\right| > C$ for each $i \neq j$, $t \geq 0$. Any finite trajectory (in the sense of the above definition) is obviously also regular, but the converse is generally not true.

Next, we consider two different cases: $p \geq 3$ and $1 < p < 3$.

**Theorem A.1** *Let the parameter* $p \geq 3$. *Then on any* finite or *regular trajectory* $[\boldsymbol{r}(t), \boldsymbol{v}(t)] \in \Omega$ *the total energy of the system* $E \geq 0$.

Proof. Let us assume the opposite, namely: on some finite or regular trajectory the total energy of the system $E < 0$. The second derivative of the function $g = g[\boldsymbol{r}(t)] = \sum_i m_i |\boldsymbol{r}_i(t)|^2$ on any finite or regular trajectory has the same form as in the case when $p = 2$, namely:

$$\ddot{g} = 2\sum_i m_i |\dot{\boldsymbol{r}}_i|^2 + 2\sum_i m_i (\ddot{\boldsymbol{r}}_i, \boldsymbol{r}_i) \tag{A.6}$$

And on this trajectory the function $\sum_i m_i(\ddot{\boldsymbol{r}}_i, \boldsymbol{r}_i)$, taking into account relations (A.2) and differential equations (A.3), A.4), satisfy the following equalities:

$$\sum_i m_i(\ddot{\boldsymbol{r}}_i, \boldsymbol{r}_i) = \sum_i \sum_{j \neq i} \frac{\gamma_p \, m_i \, m_j}{|\boldsymbol{L_{ij}}|^{1+p}} \left( \boldsymbol{r}_i, \boldsymbol{L_{ij}} \right) =$$

$$= \sum_i \sum_{j>i} \frac{\gamma_p \, m_i \, m_j}{|\boldsymbol{L_{ij}}|^{1+p}} \left( \boldsymbol{r}_i \boldsymbol{L_{ij}} + \boldsymbol{r}_j \boldsymbol{L_{ji}} \right)$$

From this, taking into account the equalities $\boldsymbol{L_{ij}} = -\boldsymbol{L_{ji}}$ and formula (A.5) for the potential function $f(\boldsymbol{r})$, we obtain the following relations on this trajectory:

$$\sum_i m_i(\ddot{\boldsymbol{r}}_i, \boldsymbol{r}_i) = \sum_i \sum_{j>i} \frac{\gamma_p \, m_i \, m_j}{|\boldsymbol{L_{ij}}|^{1+p}} \left( \boldsymbol{L_{ij}}, \boldsymbol{L_{ji}} \right) =$$

$$= -\sum_i \sum_{j>i} \frac{\gamma_p \, m_i \, m_j}{|\boldsymbol{L_{ij}}|^{p-1}} = -f[\boldsymbol{r}(t)]\,(p-1) \qquad (A.7)$$

Then from (A.6), (A.7) it follows that on this trajectory the following equality is fulfilled:

$$\ddot{g}(t) = 4E + 2(3-p)f[\boldsymbol{r}(t)] \quad \text{for each} \quad t \geq 0, \qquad \text{(A.8)}$$

where $E$ is a constant and all functions in (A.8) are continuous in $t \geq 0$ by virtue of the above definition of a finite or regular trajectory. Due to equality (A.8) and the above assumption that $E < 0$, the function $g(t)$ is strictly convex upward in $t \geq 0$ and the function $\ddot{g}(t) \leq 4E < 0$ for each $t \geq 0$. Whence it follows that the function $g(t)$ is negative for sufficiently large t, which is impossible by virtue of its definition, which means a contradiction. Thus, the inequality $E \geq 0$ is valid, which proves Theorem A.1. ■

Next, we prove that if the parameter p is in the interval: $1 < p < 3$, then on any finite trajectory the total energy of the system $E < 0$. Note that for the case when $p = 2$ this actually follows (in an approximate asymptotic sense over a large time interval) from the known theorem virial ([3], et. al.), according to which $f \sim 2T$, $E = T - f \sim (-f/2) < 0$. Below in Theorem A.2 we prove this in the exact sense for the more general model considered here and for any $1 < p < 3$.

**Theorem A.2** *Let the parameter* $1 < p < 3$. *Then on any finite trajectory* $[\boldsymbol{r}(t), \boldsymbol{v}(t)] \in \Omega$ *the total energy of the system* $E < 0$.

*Proof.* Let us assume the opposite, namely: on some finite trajectory the total energy of the system $E \geq 0$. Repeating the proof of Theorem A.1, we similarly obtain that on this trajectory a relation similar to (A.8) holds:

$$\ddot{g}(t) = 4E + 2(3-p)f[\boldsymbol{r}(t)] \quad \text{for each} \quad t \geq 0, \qquad \text{(A.9)}$$

where $E$ is a constant. Then the function $g(t)$ is strictly convex downwards in $t \geq 0$ due to the equality (A.9) and the above assumption that $E \geq 0$. Note also that on any finite trajectory the potential function of the system is bounded from below by some positive constant: $f[\boldsymbol{r}(t)] \geq B > 0$ for each $t \geq 0$ due to the inequalities $|\boldsymbol{r}_j(t) - \boldsymbol{r}_i(t)| < C < \infty$ for each $i \neq j$, $t \geq 0$. It follows from this, taking into account the equality (A.9), that the function $\ddot{g}(t)$ satisfies the inequality: $\ddot{g}(t) \geq 4E + 2(3-p)\mathrm{B} > 0$ for each $t \geq 0$, and $g(t) \to \infty$ as $t \to \infty$, i.e. this trajectory is not finite, which contradicts to the above condition of finiteness of this trajectory. Thus, the inequality $E < 0$ is valid, which proves Theorem A.2. ■

**Remark A.1** As can be seen from the previous proof, Theorem A.2 (unlike the previous Theorem A.1) is valid only for finite trajectories.

The inequality $E = T - f \geq 0$ on the trajectory of the system means that its kinetic energy $T$ exceeds the potential function $f$ coinciding with the energy required to destroy the system, i.e. to transfer all its particles (at zero initial velocities) to an infinite distance from each other. That is, the inequality $E \geq 0$ on the trajectory of the system corresponds to the infinity of this trajectory ([1] – [3] et. al.), and the previous theorems A.1 and A.2 mean that the system can exist as a single whole in some compact region of space only if the parameter $p$ is in the interval: $1 < p < 3$. And the value of the parameter $p = 2$, for which the force of attraction between particles is inversely proportional the square of the distance between them, is in the middle of this interval.

## Conclusion

Thus, the above-considered conditional minima and local extrema for the main system characteristics: total energy, size, kinetic and potential energy can be achieved only on flat (in general, not necessarily finite) trajectories of the system. In particular, along any trajectory of the system's motion in space, at any moment in time, the size of the system cannot be less than a certain (in general, unknown) lower bound which depends on the system kinetic energy at this moment and on the number N of particles and their masses. And we determine this lower bound as the solution to a nonlinear programming problem of dimension 6N for finding a conditional minimum of the size of the system on its phase space, where the system kinetic energy at a given moment of time is one of the problem's input parameters. According to the above Theorem 3.1, this lower bound can be achieved at least at one point of the trajectory only on flat, not necessarily finite, trajectories of the system's motion in space (for any number of particles and their masses). And under the conditions found above, which limit the values of the integrals of motion: angular momentum and total energy of the system these trajectories are flat finite trajectories on which the flat structure of the system is preserved in time under the restrictions on the values of the system's integrals of motion. A similar result is also hold, namely: on these same flat finite trajectories the minimum possible values of the kinetic energy and potential function of the system are achieved at the given current value of the system size. And these minimal system characteristics can take only the discrete values found above on a discrete set of the above flat finite trajectories.

Igor Pavlov, Department of Mathematics, Bauman Moscow State Technical University, ul. Baumanskaya 2-ya, 5, Moscow, 1005005, Russian Federation.

E-mail: igvalpavlov@gmail.com